\documentclass[12pt]{article}
\usepackage{amsfonts}
\usepackage{amssymb}
\usepackage{psfig}
\newcommand{\beq}{\begin{equation}}
\newcommand{\eeq}{\end{equation}}
\newcommand{\beqn}{\begin{eqnarray}}
\newcommand{\eeqn}{\end{eqnarray}}
\newcommand{\bea}[1]{\beq\begin{array}{#1}}
\newcommand{\eea}{\end{array}\eeq}

\newcommand{\Tr}[1]{\;{1\over #1}\mathop{\rm Tr}}
\newcommand{\tr}{\mathop{\rm Tr}}

\newcommand{\Texp}{\mbox{T}\!\exp}

\newcommand{\ket}[1]{|\,#1\,\rangle}
\newcommand{\bra}[1]{\langle\,#1\,|}
\newcommand{\braket}[2]{\langle\,#1\,|\,#2\,\rangle}

\newcommand{\diff}{\partial}
\newcommand{\cC}{{\cal C}}

\newcommand{\NP}[3]{{\it Nucl. Phys. }{\bf #1} (#2) #3}

\newcommand{\PL}[3]{{\it Phys. Lett. }{\bf #1} (#2) #3}
\newcommand{\PRL}[3]{{\it Phys. Rev. Lett. }{\bf #1} (#2) #3}

\newcommand{\PR}[3]{{\it Phys. Rev. }{\bf #1} (#2) #3}
\newcommand{\JL}[3]{{\it JETP Lett. }{\bf #1} (#2) #3}
\newcommand{\CMP}[3]{{\it Comm. Math. Phys. }{\bf #1} (#2) #3}

\newcommand{\PTP}[3]{{\it Prog. Theor. Phys. }{\bf #1} (#2) #3}

\begin{document}
\date{}
\title{On the non-Abelian Stokes theorem \\
for SU(2) gauge fields
\vskip-40mm
\rightline{\small ITEP-LAT/2003-21}
\vskip-4mm
\rightline{\small KANAZAWA-03-27}
\vskip 30mm
}
\author{F.V.~Gubarev\\
{\small\it ITEP, B.Cheremushkinskaya 25, Moscow, 117259, Russia}\\
{\small\it ITP, Kanazawa University, Kanazawa 920-1192, Japan}\\
}

\maketitle
\thispagestyle{empty}
\setcounter{page}{0}
\begin{abstract}
\noindent
We derive a version of non-Abelian Stokes theorem for SU(2) gauge fields
in which neither additional integration nor surface ordering are required.
The path ordering is eliminated by introducing the instantaneous color
orientation of the flux.
We also derive the non-Abelian Stokes theorem on the lattice and discuss
various terms contributing to the trace of the Wilson loop.
\end{abstract}

\newpage
%====================================================================

\subsection*{Introduction}
\noindent
The usual Abelian Stokes theorem relates the integral along closed
curve $\cC$ bounding some surface $S_\cC$ and an integral defined
on this surface. This version of the Stokes theorem is most relevant
in physical applications since it allows to express the holonomy of
gauge non-invariant electromagnetic potential via physically
observable magnetic flux.
The invention of non-Abelian gauge theories necessitated a non-Abelian
generalization of the Stokes theorem. Nowadays, there are quite a few
formulations of non-Abelian Stokes theorem (NAST) available
(for review see, e.g., Ref.~\cite{Broda:2000id} and references therein).
Generically there exist two principal approaches, the
operator~\cite{Halpern:1978ik}---\cite{Shevchenko:1998uw}
and the path-integral~\cite{Diakonov:fc}---\cite{Zubkov:2002ei} one.

It is worth to mention that the central issue of any formulation of NAST
is how to make sense of the path ordering prescription
inherent to the non-Abelian holonomy (Wilson loop).
In this respect neither operator nor path-integral approaches are
helpful for concrete calculation of the Wilson loop.
Indeed, while in the former case the path ordering is traded for much
more complicated surface ordering prescription, in the latter case
the additional path integral over auxiliary variables is introduced
which cannot be calculated even approximately.

In this paper we derive a new version of non-Abelian Stokes theorem
focusing exclusively on $SU(2)$ valued Wilson loops in the fundamental
representation. Moreover, the central object of our discussion
is the phase of the Wilson loop $\varphi_w$. It determines
the Wilson loop trace, $1/2\tr W = \cos\varphi_w$, 
which is the only gauge invariant quantity associated with gauge holonomy.

The basic idea is to introduce the instantaneous color orientation
of the chromo-magnetic flux piercing the loop. Evidently this color
orientation remains unknown until the Wilson loop is calculated by some
other means. Nevertheless, it allows us to avoid the path ordering
and represent the Wilson loop phase as an ordinary integral to which
Abelian Stokes theorem applies. Furthermore, we relate the resulting
surface integral with properties of gauge fields on this surface.

As we noted above, in order to get explicitly the color orientation of
the flux one has to calculate the Wilson loop first. In this respect
our formulation is well suited for the lattice where the gauge holonomy
is to be calculated numerically. We derive the lattice version
of the non-Abelian Stokes theorem. Finally, we discuss the physical meaning
and origin of various terms contributing to the trace of the Wilson loop
and present the results of our qualitative numerical simulations.

%====================================================================
\subsection*{Non-Abelian Stokes theorem in the continuum limit}
\noindent
Consider Wilson loop operator in the fundamental representation, $W(T)$,
evaluated on a closed contour $\cC = \{ x_\mu(t), t\in [0;T], x_\mu(0) = x_\mu(T)\}$,
which is parameterized by differentiable functions $x_\mu(t)$.
By definition the operator $W(T)$ provides a solution to
the first-order differential equation
\beq
\label{defEQ}
\bra{\psi(t)}\, \left( i \diff_t ~+~ A \right) ~=~ 0\,,
\eeq
\beq
\label{prop}
\bra{\psi(t)} ~=~ \bra{\psi(0)} \, W(t)\,,
\eeq
\beq
W(t)~=~ \Texp\{ i \int_0^t A(\tau) d\tau \}\,,
\eeq
where $A = 1/2 \sigma^a A^a_\mu[x(t)]\,\dot{x}_\mu(t)$ is the tangential component of the
gauge potential, $\sigma^a$ are the Pauli matrices, differential is taken with respect
to the parameter $t$ and $\bra{\psi}$ is a vector in the spin-1/2 irreducible
representation space (IRR) of $SU(2)$ group.
Since Eq.~(\ref{defEQ}) is nothing else but the time-dependent Schr\"odinger equation,
the Wilson loop $W(t)$ can be interpreted as a quantum mechanical evolution operator
with the time-dependent Hamiltonian $H = - A(t)$. Moreover, the corresponding state space
coincides with spin-1/2 IRR, in which a convenient basis  is provided by generalized
(spin) coherent states~\cite{Perelomov:1971bd}
(see, e.g., Ref.~\cite{Perelomov:yw}---\cite{Zhang:1999is} for review).
The spin coherent states $\{\bra{\vec{n}}\}$ are parameterized
by a set of unit three-dimensional vectors $\vec{n}$, $\vec{n}^2 = 1$ and
in this basis arbitrary state $\bra{\psi}$ has a unique representation
\beq
\label{coherentRep}
\bra{\psi} ~=~ e^{ i\varphi} \,\, \bra{\vec{n}}\,.
\eeq
The action of $SU(2)$ operator $g$ reads
\beq
\bra{\vec{n}}\,\,g ~=~ e^{ i\varphi} \,\, \bra{\vec{n}_g}\,,
\eeq
where the phase factor $\varphi$ depends on both $\vec{n}$ and $g$,
$\varphi = \varphi(g,\vec{n})$.
Therefore, in the basis of coherent states Eq.~(\ref{prop}) becomes
\beq
\label{constant}
\bra{\psi(t)} ~=~ e^{ i\varphi(t)} \,\, \bra{\vec{n}(t)} ~=~ \bra{\vec{n}(0)} \, W(t)\,,
\eeq
where without loss of generality we have taken
$\varphi(0) = 0$ or equivalently $\bra{\psi(0)} = \bra{\vec{n}(0)}$.

Eq.~(\ref{defEQ}) imposes no restrictions on the initial vector $\bra{\vec{n}(0)}$,
which therefore can be taken arbitrarily. However, there exists a distinguished initial state
which is of particular importance for the discussion below. Namely, let us take
$\bra{\vec{n}(0)}$ to be the eigenstate of the full evolution operator
\beq
\label{eigen}
\bra{\vec{n}(0)}\,\, W(T)  ~=~  e^{ i\varphi(T)} \,\, \bra{\vec{n}(0)}\,.
\eeq
Note that generically $1/2\tr W(T)\ne\pm 1$ and we assume this from now on.
The gauge invariant trace of the Wilson loop is given by $1/2\tr W(T) = \cos\varphi(T)$.
On the other hand, one gets from (\ref{defEQ},\ref{coherentRep}) the following
equation for the Wilson loop phase factor~\cite{Gubarev:2000qg}
\beq
\label{Phase0}
\varphi(T) ~=~ \int_0^T \, \Big(
\bra{\vec{n}}\,A\,\ket{\vec{n}} ~-~ i \bra{\vec{n}}\,\diff_t\,\ket{\vec{n}} \Big)\, dt\,.
\eeq
Using standard properties of the spin coherent states~\cite{Perelomov:yw}---\cite{Zhang:1999is}
one can represent (\ref{Phase0}) in vector-like notations
\beq
\label{Phase}
\Tr{2} W(T) ~=~ \cos\left[\,\, \frac{1}{2}\int_C \vec{n}\vec{A}\,dt ~+~
\frac{1}{4} \int_{S_C} \vec{n} \cdot [ \partial_\mu \vec{n} \times \partial_\nu \vec{n}]
\,\,d^2\sigma^{\mu\nu} \,\,\right]\,,
\eeq
where the vector field $\vec{n}(t)$ has been smoothly extended from the
contour $\cC$ into an arbitrary surface $S_\cC$ bounded by $\cC$.
Note that Eq.~(\ref{Phase}) may be identically rewritten in the gauge invariant form
\beq
\label{Phase2}
\Tr{2} W(T) ~=~ \cos\left[\,\, \frac{1}{4} \int_{S_C}\,
\left\{ \vec{n} \vec{F}_{\mu\nu} ~+~ 
\vec{n} \cdot [ D_\mu \vec{n} \times D_\nu \vec{n}]\,
\right\} \,d^2\sigma^{\mu\nu} \,\,\right]\,,
\eeq
where $D^{ab}_\mu = \delta^{ab}\diff_\mu - \varepsilon^{acb} A^c_\mu$
is the covariant derivative and
$F^a_{\mu\nu} = \diff_\mu A^a_\nu - \diff_\nu A^a_\mu -  \varepsilon^{abc} A^b_\mu A^c_\nu$
is the non-Abelian field strength.

Let us emphasize that  Eqs.~(\ref{Phase},\ref{Phase2})
cannot be used to actually calculate the Wilson loop
since construction of the evolving states $\vec{n}(t)$ requires knowledge
of the Wilson loop itself. Nevertheless, Eq.~(\ref{Phase2}) might be
relevant for theoretical considerations since it represents the phase factor
$\varphi(T)$ as an integral of the field strength introduced first by
't~Hooft~\cite{'tHooft:1974qc} and Polyakov~\cite{Polyakov:ek} in connection with monopoles.
Eq.~(\ref{Phase2}) is the non-Abelian Stokes theorem (NAST) to be discussed
in more detail below.

By construction the vector field $\vec{n}(t)$ is covariantly constant
along the contour $\cC$, $\dot{x}_\mu \, D_\mu \vec{n} = 0$.
Eq.~(\ref{constant}) implies also that for any $t \in [0;T]$
the state $\bra{\vec{n}(t)}$ is an eigenstate of the Wilson loop calculated
on $\cC$ starting from the point $x_\mu(t)$. In other words, $\bra{\vec{n}(t)}$
is an eigenstate of $W^+(t) \, W(T) \, W(t)$.
The interpretation of $\vec{n}(t)$ is then straightforward, it is
the instantaneous color orientation of the flux piercing the loop $\cC$.
Since the contour considered is not infinitesimal the color direction
of the flux is different at various points on $\cC$.

The assignment of the vector field $\vec{n}(t)$ to a given closed path
$\cC$ is unfortunately not unique. The ambiguity comes from Eq.~(\ref{eigen})
which possesses two solutions with opposite sign of $\vec{n}$
\beq
\label{eigen2}
\bra{\,\pm\,\vec{n}(0)}\,\, W(T) ~=~ e^{\,\pm\,i\varphi(T)} \,\,
\bra{\,\pm\,\vec{n}(0)}\,.
\eeq
However, this sign ambiguity is only global: if $\bra{\vec{n}(0)}$
is an eigenstate with $\varphi(T) > 0$, then for any $t\in[0;T]$ $\bra{\vec{n}(t)}$
is an eigenstate of $W^+(t) \, W(T) \, W(t)$ with the same positive phase. Therefore there is
only a global freedom to change $\vec{n}(t)\to - \vec{n}(t)$ for all $t$ simultaneously.

Consider the infinitesimal closed contour $\delta \cC_x$ located in $(\mu,\nu)$ plane 
at point $x$, which bounds the elementary surface element $\delta\sigma^{\mu\nu}_x$.
Since in this case
$W(T) \equiv W(\delta\cC_x) = 1 + \frac{i}{2}\vec{\sigma} \vec{F}_{\mu\nu}\delta\sigma^{\mu\nu}_x + o(\delta\sigma)$
the eigenvalue problem (\ref{eigen},\ref{eigen2}) could easily be solved
\beq
\label{eigen-infinitesimal}
\vec{n}^{(\mu\nu)}_x ~=~ \pm\frac{\vec{F}\delta\sigma}{|\vec{F}\delta\sigma|}\,,
\qquad
\varphi(\delta\cC_x) ~=~ \pm |\vec{F}\delta\sigma| ~=~
\pm \sqrt{\left(\vec{F}\delta\sigma\right)^2}\,,
\eeq
where we have explicitly indicated the $(\mu\nu)$ dependence of the eigenvector and
evident Lorenz indices have been suppressed.
Therefore the infinitesimal version of non-Abelian Stokes theorem (\ref{Phase2}) is
(no summation over $\mu$, $\nu$)
\beq
\label{infinitesimal}
\Tr{2} W(\delta\cC_x) \approx \cos\left[\,\,
\frac{1}{2} \vec{n}^{(\mu\nu)}_x \vec{F}_{\mu\nu}(x) \delta\sigma^{\mu\nu}_x +
o( \delta\sigma )\,\,\right] \approx
1 - \frac{1}{8} \left(\vec{F}_{\mu\nu}\delta\sigma^{\mu\nu}_x\right)^2 + o(\delta\sigma^2)\,,
\eeq
where we have used the covariant constancy of $\vec{n}$ on $\delta\cC_x$.

It is amusing to note that Eqs.~(\ref{Phase}, \ref{Phase2}) look similar
to the non-Abelian Stokes theorem of Ref.~\cite{Diakonov:fc}
apart from the absence of the path integral over $\vec{n}(t)$ in Eq.~(\ref{Phase2}).
We conclude therefore that the path integral of Ref.~\cite{Diakonov:fc}
is exactly saturated by the two particular trajectories $\pm\,\vec{n}(t)$. It is worth
mentioning however that the construction of $\vec{n}(t)$,
Eq.~(\ref{constant}), makes no reference to the classical equations of motion
and therefore Eqs.~(\ref{Phase},\ref{Phase2}) do not correspond in general
to any semi-classical approximation.

Notice that Eq.~(\ref{Phase2}) is still not uniquely defined. The point
is that the vector field $\vec{n}(t)$ may be extended arbitrarily from $\cC$ to $S_\cC$.
The only requirement is that the extension 
$\cC \ni \vec{n}(t) \to \vec{n}(\sigma)\in S_\cC$ must be continuous and
the distribution $\vec{n}(\sigma)$ must agree with $\vec{n}(t)$
at the boundary $\delta S_\cC = \cC$. On the other hand, Eq.~(\ref{Phase2})
is applicable to any closed contour, in particular to any infinitesimal
area element of $S_\cC$. Therefore, at every point $x(\sigma) \in S_\cC$
we have a naturally defined direction $\vec{n}(\sigma)$ which can be used
to make Eq.~(\ref{Phase2}) unambiguous. Then the only remaining problem
is the choice of sign of $\vec{n}(\sigma)$ since at every
point $x(\sigma) \in S_\cC$ considered separately there is no distinction
between $\pm \vec{n}(\sigma)$. Here the continuity requirement for the
extension $\cC\ni\vec{n}(t)\to \vec{n}(\sigma)\in S_\cC$ becomes
crucial. Indeed, since the surface $S_\cC$ is assumed to be regular
(e.g., smooth and without self-intersections) the field
$\vec{F}\delta\sigma$ is continuous on $S_\cC$ and therefore
Eq.~(\ref{eigen-infinitesimal}) allows to define $\vec{n}(\sigma)$ continuously as well
(these arguments may fail for exceptional configurations
which are not generic and which we don't consider for that reason).

The sign ambiguity in the definition of $\vec{n}(\sigma)$
is reminiscent to the model of Ref.~\cite{Schwarz:ec,Schwarz:zt} (Alice electrodynamics).
Indeed, the key feature of Alice electrodynamics is that
the $U(1)$ generator ($\vec{n}(\sigma)$ in our case) is known only up to the sign.
Moreover, the definition of the central object of the model, Alice string,
is based on the continuity arguments similar to the above reasoning.
However, the relevance of Alice electrodynamics to the $SU(2)$ gauge theory
is still unclear and we will not dwell on this issue.

In the next Section we consider the non-Abelian Stokes theorem
on the lattice. As a byproduct we also illustrate the appearance
of various terms in Eq.~(\ref{Phase2}).

%====================================================================
\subsection*{Non-Abelian Stokes theorem on the lattice}
\noindent
The derivation of lattice NAST begins from consideration of
fundamental representation Wilson loop on the lattice
\beq
\label{loop}
W ~=~ \prod\limits^{N-1}_{i=0}\, U_i ~=~ U_0 \cdot U_1 \cdot ... \cdot U_{N-1}\,,
\eeq
where $U_i\in SU(2)$ are the link matrices parameterized as
\beq
\label{param}
U_i = \cos\upsilon_i + i \, \sin\upsilon_i \,\, \vec{\sigma} \, \vec{u}_{i,+} =
      \cos\upsilon_i - i \, \sin\upsilon_i \,\, \vec{\sigma} \, \vec{u}_{i,-} \,,
\qquad
\vec{u}^2_{i,\pm} = 1\,,
\eeq
and the corresponding lattice path piercing the sites
$s_0, s_1, ..., s_{N-1}$ is assumed to be closed and non self-intersecting.
There is an analogous parameterization of the Wilson loop
\beq
W = \cos\varphi_w + i \, \sin\varphi_w \,\, \vec{\sigma} \, \vec{w}_+
= \cos\varphi_w - i \, \sin\varphi_w \,\, \vec{\sigma} \, \vec{w}_-\,,
\qquad
\vec{w}^2_\pm = 1\,,
\eeq
and therefore the state to be ascribed to the site $s_0$ is 
$\bra{\vec{w}_+} \equiv \bra{\vec{w}_+(s_0)}$
\beq
\label{eigen0}
\bra{\vec{w}_+(s_0)}\,W = e^{ i \varphi_w}\,\,\bra{\vec{w}_+(s_0)}\,.
\eeq
Note that one could equally take the eigenstate
$\bra{\vec{w}_-(s_0)}$ instead (see discussion in the previous Section).
As far as only a single Wilson loop is concerned, there are no much difference
between the two choices and it is sufficient to take either of them.

Starting from $\bra{\vec{w}_+(s_0)}$ one constructs the corresponding eigenstates
$\bra{\vec{w}_+(s_i)}$, $i=1,...,N-1$ in all other sites $s_1,...,s_{N-1}$
using Eqs.~(\ref{coherentRep},\ref{constant})
\beq
\label{chain}
\bra{\vec{w}_+(s_i)}\,\, U_i ~=~ e^{i\gamma_i} \, \bra{\vec{w}_+(s_{i+1})}\,,
\qquad
i = 0, ... , N-1\,.
\eeq
Since the initial state was taken to be an eigenstate of $W$, the chain
(\ref{chain}) is closed, $s_N\equiv s_0$,
$\bra{\vec{w}_+(s_{N})} \equiv \bra{\vec{w}_+(s_0)}$
and one gets the following relation between the Wilson loop phase $\varphi_w$
and the phases $\gamma_i$ coming from links $U_i$~\cite{Gubarev:2000qg}
\beq
\label{NAST-1}
\varphi_w ~=~ \sum\limits^{N-1}_{i=0} \, \gamma_i\,.
\eeq
We are in haste to add that, strictly speaking, Eq.~(\ref{NAST-1}) is not entirely
correct. The point is that the left hand side is always bounded,
$|\varphi_w| \leq \pi$ while the sum on the right can take values
outside the interval $[-\pi;\pi]$. To be precise, Eq.~(\ref{NAST-1}) should
express the equality of the phase factors $e^{i\varphi_w}$, not the
angles $\varphi_w$ by themselves. Therefore, there are terms $2\pi k$, $k\in Z$
missing in Eq.~(\ref{NAST-1}). From now on in all the equations like (\ref{NAST-1})
the $\mathrm{~mod~}2\pi$ operation is always assumed and will not be indicated
explicitly.

In order to make one step further we need to consider in more detail Eq.~(\ref{chain}).
Consider an $SU(2)$ operator $U$ which upon acting on some initial state
$\bra{\vec{n}_1}$ brings it to another state $\bra{\vec{n}_2}$
\beq
\label{action}
\bra{\vec{n}_1}\,\, U ~=~ e^{i\gamma} \, \bra{\vec{n}_2}\,, \qquad
U = \cos\upsilon + i \, \sin\upsilon \,\, \vec{\sigma} \, \vec{u}_+\,,
\eeq
where $\gamma$ is an additional phase which depends
on both $\bra{\vec{n}_1}$ and $U$. There are two eigenstates 
$\bra{\vec{u}_{\pm}}$ of the operator $U$ which form a complete basis
in the spin-1/2 IRR
\beq
\label{unity}
\bra{\vec{u}_{\pm}} \,\, U ~=~ e^{ \pm i\upsilon } \, \bra{\vec{u}_{\pm}}\,,
\qquad
1 ~=~ \ket{\vec{u}_+}\bra{\vec{u}_+} ~+~ \ket{\vec{u}_-}\bra{\vec{u}_-}\,.
\eeq
Using the resolution of unity (\ref{unity}) in (\ref{action}) one gets
\beq
e^{i\upsilon}\,\braket{\vec{n}_1}{\vec{u}_+} \,\, \bra{\vec{u}_+} ~+~
e^{-i\upsilon}\,\braket{\vec{n}_1}{\vec{u}_-} \,\, \bra{\vec{u}_-} ~=~
e^{i\gamma} \, \bra{\vec{n}_2}\,,
\eeq
\beq
\label{gamma}
\gamma ~=~ \upsilon ~+~ \Omega_0(\vec{n}_1, \vec{u}_+, \vec{n}_2) ~=~
- \upsilon ~+~ \Omega_0(\vec{n}_1, \vec{u}_-, \vec{n}_2)\,,
\eeq
where $\Omega_0(\vec{n}_1, ... , \vec{n}_N)$ is the oriented area of the spherical polygon
on unit two-dimensional sphere $S^2$ with corners at the north pole of $S^2$
and at $\vec{n}_1, ... , \vec{n}_N$ (in that order). In deriving Eq.~(\ref{gamma})
standard properties of the spin coherent states have been used
(see, e.g., Ref.~\cite{Perelomov:yw}---\cite{Zhang:1999is})
$$
\braket{\vec{n}_1}{\vec{n}_2}\,\slash\, |\braket{\vec{n}_1}{\vec{n}_2}| ~=~
e^{ i \Omega_0(\vec{n}_1,\vec{n}_2) }\,.
$$
Note that in our normalization $\mathrm{Area}(S^2) = 2\pi$. In particular, there
is no additional $1/2$ factor in front of $\Omega_0$.

{}From very general arguments one expects that Eq.~(\ref{action}) has an interpretation
of ordinary rotation of the vector $\vec{n}_1$ around the axis $\vec{u}_+$ by
the angle $\upsilon$. However, as far as only initial and final states are
taken into account the rotation operator remains in fact undetermined.
Indeed, there are infinitely
many rotations which connect two given states.
On the other hand, among various $SU(2)$ operators
there is a distinguished unique $G_{n_1\to n_2}$ which
describes the motion $\vec{n}_1\to \vec{n}_2$ along the shortest geodesic line
connecting $\vec{n}_1$, $\vec{n}_2$
\beq
G_{n_1\to n_2} ~=~ (\vec{n}_1 \, \vec{n}_2 ) +
i \vec{\sigma} \cdot [\vec{n}_1 \times \vec{n}_2]\,,
\qquad
\bra{\vec{n}_1}\,\, G_{n_1\to n_2} = e^{i\Omega_0(\vec{n}_1,\vec{n}_2)} \, \bra{\vec{n}_2}\,.
\eeq
The physical relevance of geodesic curves is widely known.
The importance of the geodesic matrices $G_{n_1\to n_2}$ in the present
context comes from the consideration of the following diagram
$$
\begin{array}{rcl}
\vec{n}_1 & \stackrel{U}{\longrightarrow} & \vec{n}_2 \\
\left. G_{m_1\to n_1} \rule{0mm}{5mm} \right\uparrow
& & 
\left\uparrow \rule{0mm}{5mm} G_{m_2\to n_2} \right.
\\
\vec{m}_1 & \stackrel{U}{\longrightarrow} & \vec{m}_2 \\
\end{array}
$$
Here the same operator $U$ corresponds to the rotations $\vec{n}_1\to \vec{n}_2$
and $\vec{m}_1\to \vec{m}_2$. The diagram is closed by two geodesic matrices
$G_{m_1\to n_1}$ and $G_{m_2\to n_2}$. From the analysis of the diagram
one obtains the relation
\beq
\label{eq1}
G_{m_1\to n_1} \, U ~=~ U \, G_{m_2\to n_2}\,,
\eeq
which follows from the fact that adjoint $SU(2)$ action is equivalent
to $SO(3)$ rotation. Therefore the matrix $U G_{m_2\to n_2} U^+$
is again geodesic with initial and final states being $\vec{m}_1$ and $\vec{n}_1$
and thus equals to $G_{m_1\to n_1}$. Eq.~(\ref{gamma}) allows to rewrite 
(\ref{eq1}) in the form
\beq
\label{eq2}
\Omega_0(\vec{n}_1, \vec{u}_\pm, \vec{n}_2) +
\Omega_0(\vec{m}_2, \vec{u}_\pm, \vec{m}_1) =
\Omega_0(\vec{m}_2, \vec{n}_2 ) +
\Omega_0(\vec{n}_1, \vec{m}_1)\,.
\eeq

Eqs.~(\ref{NAST-1},\ref{gamma},\ref{eq2}) are sufficient to derive the NAST on the lattice.
We illustrate the derivation on the simplest example. Generalization to usual cubic
geometry is straightforward but technically notorious. Thus we will only discuss the final
result.

\begin{figure}[ht]
\centerline{\psfig{file=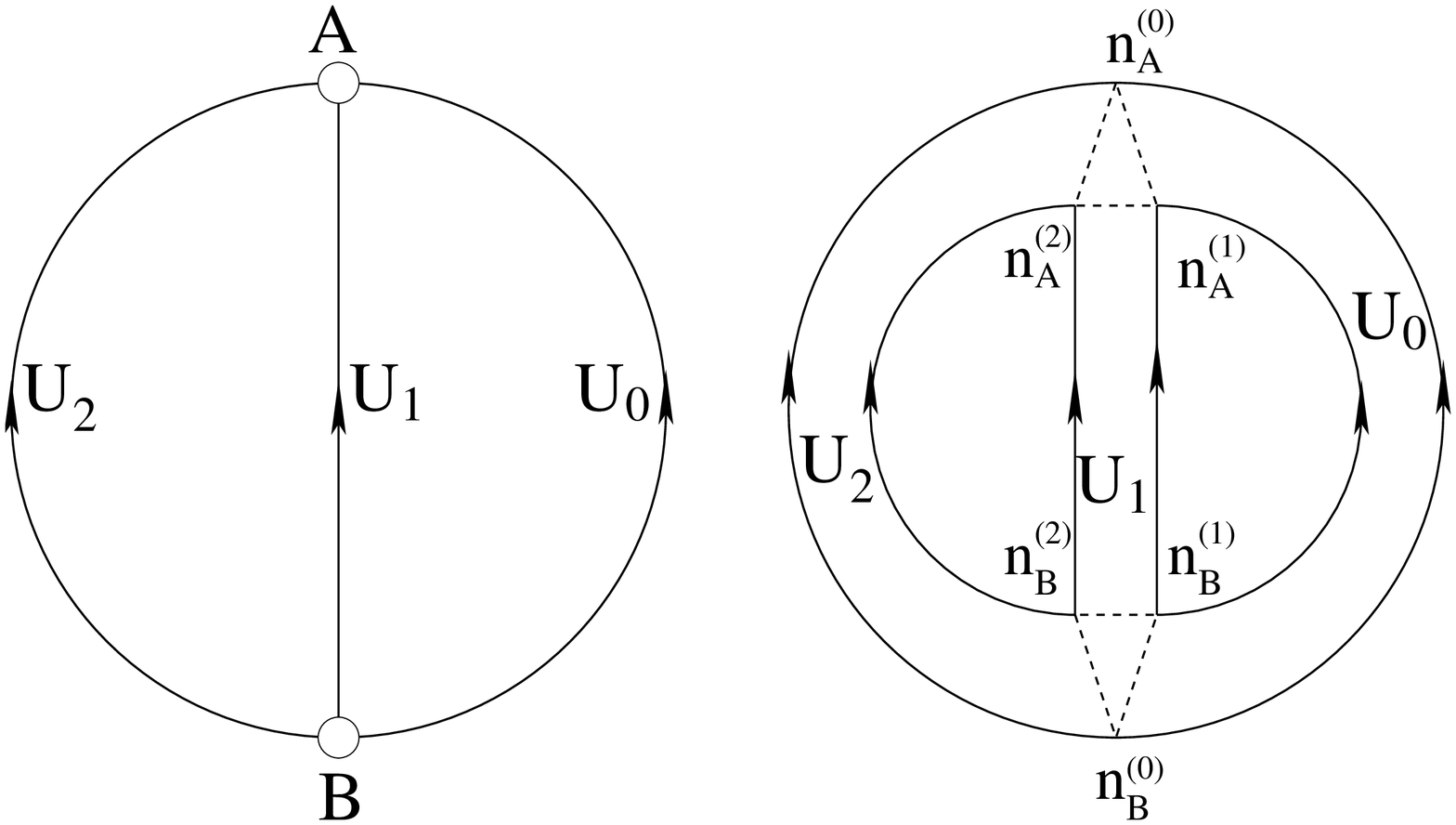,width=0.6\textwidth,silent=}}
\centerline{Fig.~1}
\end{figure}

Consider the simplest non-trivial configuration of three links, Fig.~1,
from which one could construct three different Wilson loops
$$
W_0 ~=~ U_0\,U^+_2\,, \quad
W_1 ~=~ U_0\,U^+_1\,, \quad
W_2 ~=~ U_1\,U^+_2\,,
$$
\beq
W_0 ~=~ W_1 \,\, W_2\,.
\eeq
Let $\varphi_i$ is the phase angle of the corresponding Wilson loop,
$1/2\tr W_i = \cos\varphi_i$. 
Applying the procedure of the previous Section to each Wilson loop separately, one gets six
states, $\{ \vec{n}^{(i)}_A, \vec{n}^{(i)}_B\}$, $i = 0,1,2$, sitting at the points
A and B; pair of states with fixed $i$ is assigned to the corresponding Wilson loops $W_i$.
In particular,
\beq
\label{eigen3}
\bra{\vec{n}^{(i)}_A} \,\, W_i ~=~ e^{i\varphi_i} \, \bra{\vec{n}^{(i)}_A}\,,
\qquad
i = 0, 1, 2\,.
\eeq
Let us evaluate the phase angle $\varphi_0$. According to (\ref{NAST-1},\ref{gamma})
\beq
\varphi_0 ~=~ \upsilon_0 + \Omega_0(\vec{n}^{(0)}_A, \vec{u}_{0,+}, \vec{n}^{(0)}_B)
- \upsilon_2 + \Omega_0(\vec{n}^{(0)}_B, \vec{u}_{2,+}, \vec{n}^{(0)}_A)\,,
\eeq
where $\upsilon_2$ enters with minus sign because $U_2$ is conjugated in
the definition of $W_0$. Using Eq.~(\ref{eq2}) we can write
\beq
\label{junk1}
\begin{array}{ccr}
\varphi_0 & = &
\upsilon_0 + \Omega_0(\vec{n}^{(0)}_A, \vec{n}^{(1)}_A) +
\Omega_0(\vec{n}^{(1)}_A, \vec{u}_{0,+}, \vec{n}^{(1)}_B) +
\Omega_0(\vec{n}^{(1)}_B, \vec{n}^{(0)}_B)\,\, \\
& & \\
& &
- \upsilon_2 +  \Omega_0(\vec{n}^{(0)}_B, \vec{n}^{(2)}_B) +
\Omega_0(\vec{n}^{(2)}_B, \vec{u}_{2,+}, \vec{n}^{(2)}_A) +
\Omega_0(\vec{n}^{(2)}_A, \vec{n}^{(0)}_A)\,.
\end{array}
\eeq
The next step is to add zero in the form (see (\ref{eq1},\ref{eq2}))
\beq
0 = \Omega_0(\vec{n}^{(1)}_A, \vec{n}^{(2)}_A) +
\Omega_0(\vec{n}^{(2)}_B, \vec{n}^{(1)}_B) +
\upsilon_1 + \Omega_0(\vec{n}^{(2)}_A, \vec{u}_{1,+}, \vec{n}^{(2)}_B)
- \upsilon_1 + \Omega_0(\vec{n}^{(1)}_B, \vec{u}_{1,+}, \vec{n}^{(1)}_A)
\eeq
to Eq.~(\ref{junk1}) and collect various terms together using the relations like
\bea{c}
\varphi_1 = \upsilon_0 + \Omega_0(\vec{n}^{(1)}_A, \vec{u}_{0,+}, \vec{n}^{(1)}_B)
- \upsilon_1 + \Omega_0(\vec{n}^{(1)}_B, \vec{u}_{1,+}, \vec{n}^{(1)}_A)\,,
\\
\\
\Omega(A) = 
\Omega_0(\vec{n}^{(0)}_A, \vec{n}^{(1)}_A) +
\Omega_0(\vec{n}^{(1)}_A, \vec{n}^{(2)}_A) +
\Omega_0(\vec{n}^{(2)}_A, \vec{n}^{(0)}_A)\,,
\eea
where $\Omega(A) = \Omega(\vec{n}^{(0)}_A, \vec{n}^{(1)}_A, \vec{n}^{(2)}_A)$
is the area of spherical triangle constructed on the indicated three unit vectors at point A.
In other words,  $\Omega(A)$ is just the oriented solid angle between the triple
$\{\vec{n}^{(0)}_A, \vec{n}^{(1)}_A, \vec{n}^{(2)}_A\}$.

Therefore the final equation which relates the phase angles $\varphi_i$, $i=0,1,2$ is
\beq
\label{NAST-2}
\varphi_0 ~=~ \varphi_1 ~+~ \varphi_2 ~+~ \Omega(A) ~+~ \Omega(B)\,,
\eeq
where $\Omega(A)$, $\Omega(B)$ are the oriented solid angles between the triads
$\{\vec{n}^{(0)}_A, \vec{n}^{(1)}_A, \vec{n}^{(2)}_A\}$ and
$\{\vec{n}^{(0)}_B, \vec{n}^{(2)}_B, \vec{n}^{(1)}_B\}$.
Note the different ordering of states in $\Omega(A)$ and $\Omega(B)$, which
corresponds to counting the outgoing flux at both points A and B.

Let us emphasize that Eq.~(\ref{NAST-2}) is valid irrespectively of the particular choice
of the states $\{ \vec{n}^{(i)}_A, \vec{n}^{(i)}_B\}$ provided that the phases
$\varphi_i$ are calculated according to (\ref{NAST-1}) or (\ref{eigen3}).
It does not matter
which particular solution of (\ref{eigen3}) was taken to construct
$\{ \vec{n}^{(i)}_A, \vec{n}^{(i)}_B\}$ on each Wilson loop,
Eq.~(\ref{NAST-2}) remains formally the same with either choice.
But this means that Eq.~(\ref{NAST-2}) is ambiguous because $\varphi_i$
changes sign when $\{ \vec{n}^{(i)}_A, \vec{n}^{(i)}_B\}$ are replaced
by $\{ -\vec{n}^{(i)}_A, -\vec{n}^{(i)}_B\}$. In fact this is the same sign problem
discussed previously. We will fix it after considering the continuum limit
of Eq.~(\ref{NAST-2}).

\begin{figure}[ht]
\centerline{\psfig{file=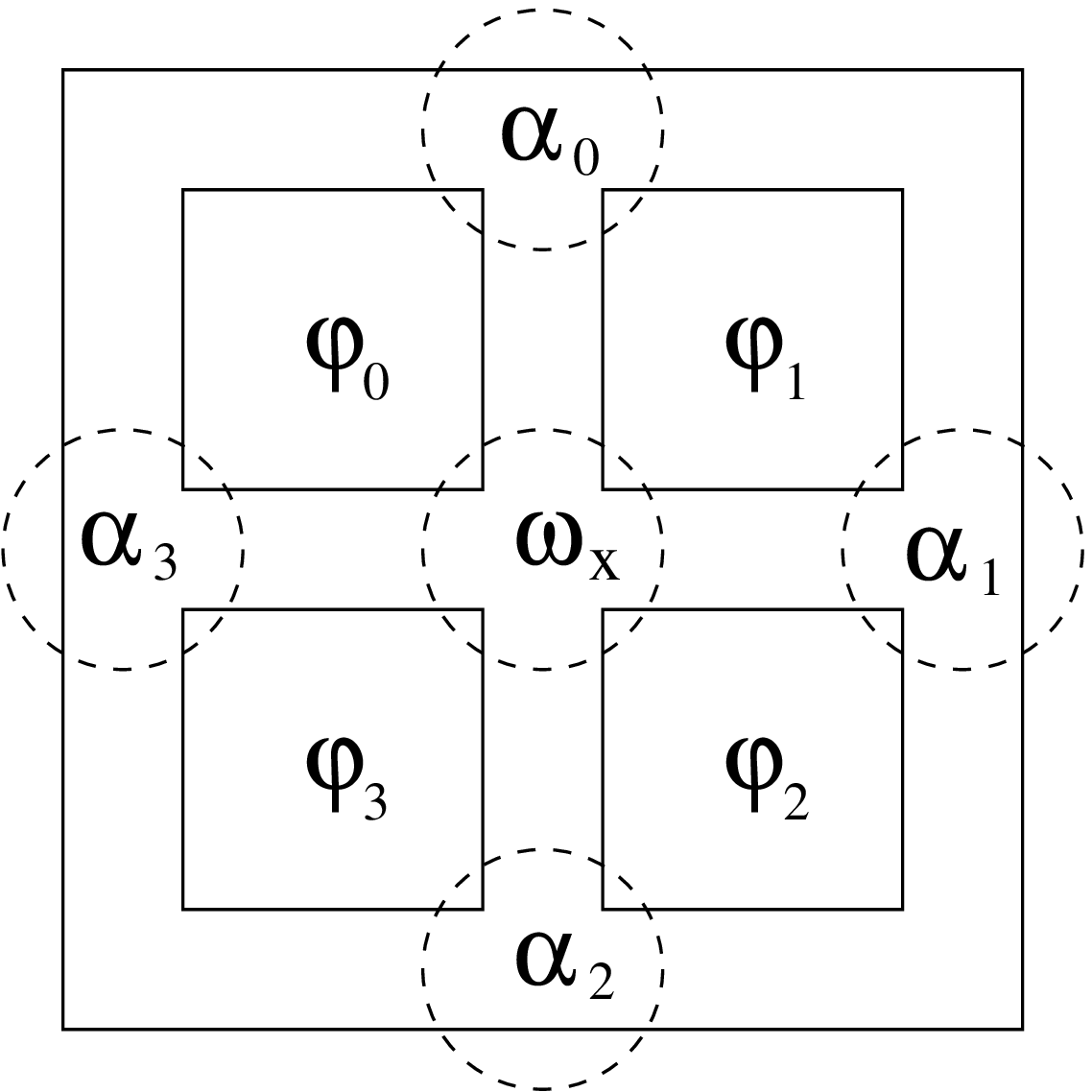,width=0.3\textwidth,silent=}}
\centerline{Fig.~2}
\end{figure}

The generalization of Eq.~(\ref{NAST-2}) to the case of usual cubic geometry is 
straightforward but technically involved. The final result
\beq
\label{NAST-3}
\varphi_w ~=~ \sum\limits_{x\in S_\cC} \varphi_x ~+~ 
\sum\limits_{x\in S_\cC} \Omega_x ~+~  \sum\limits_{x\in\cC} \alpha_x
\eeq
is illustrated on Fig.~2. It is understood that phases $\varphi_w$,
$\varphi_x$ are calculated via Eq.~(\ref{NAST-1}). $\varphi_w$ is the phase
of the large Wilson loop, $1/2\tr W(\cC) = \cos\varphi_w$, where $\cC$ is the
planar $2\times 2$ closed contour, see Fig.~2, which bounds the surface $S_\cC$.
The first term on the r.h.s. ('dynamical part') is the sum of contributions coming
from four 'internal' plaquettes belonging to $S_\cC$. In particular,
$$
\Tr{2}\, U_{p_x} ~=~ \cos\varphi_x\,, \qquad  p_x = 0, 1, 2, 3\,,
$$
where $U_{p_x}$ is the corresponding plaquette matrix. The second term ('solid angle')
comes from the points common to four different 'internal' plaquettes. We recall that
application of NAST, Eq.~(\ref{NAST-2}), requires construction of four
color vectors per plaquette situated at plaquette's corners. Therefore there are
four unit vectors at the point $x$, see Fig.~2, and $\Omega_x$ is just the oriented
solid angle between them. The third term ('perimeter contribution') is analogous
to the second one. It accounts for the difference in color direction between
the states on the nearest to the loop 'internal' plaquettes and the states on the loop
itself. Technically $\alpha_i$ is an oriented solid angle between the corresponding three vectors.

Eq.~(\ref{NAST-3}) has a simple physical interpretation. The magnitude of the total flux,
$\varphi_w$, piercing large closed contour $\cC$ is the sum of a few terms.
The first term sums up the magnitudes of elementary fluxes penetrating the surface
$S_\cC$. Since the theory is non-Abelian each elementary flux has its own color
orientation which is no less important than the flux magnitude (for
flux piercing finite contour $\cC$ the color direction of the flux varies along $\cC$).
The other terms in Eq.~(\ref{NAST-3}) take into account the difference in color
orientation of various fluxes on $S_\cC$ as well as of the total flux piercing $\cC$. 
It is worth to mention that for pure Abelian fields (or for a $SU(2)$ gauge copy of Abelian
configurations) the second and third contributions in (\ref{NAST-3}) vanish identically and
one gets the usual Abelian Stokes theorem.

Let us consider Eq.~(\ref{NAST-3}) in the limit of vanishing lattice spacing, $a\to 0$.
The contribution of the first term was in fact already calculated in
(\ref{eigen-infinitesimal},\ref{infinitesimal})
\beq
\label{dyn}
\mathrm{'dynamical~~part'~} ~=~
a^2 \, \sum\limits_{x\in S_\cC} \, \frac{1}{2} \vec{n}_x \vec{F}_{\mu\nu}(x) + o(a^2)
~\approx~ \frac{1}{4} \int_{S_C}\,\vec{n} \vec{F}_{\mu\nu} \,d^2\sigma^{\mu\nu} \,,
\eeq
where $\vec{n}_x$, $x\in S_\cC$ is given by (\ref{eigen-infinitesimal}). In order
to get the continuum limit of the second term consider the point
$x\in S_\cC$ and let $(\mu\nu)$ is a plane tangential to $S_\cC$ at $x$.
Then $\Omega_x$ is the oriented solid angle between the four vectors
\bea{l}
\vec{n}_1 ~=~ \vec{n}_x ~-~ a \cdot \left( D_\mu \vec{n}_x + D_\nu \vec{n}_x \right)\,,\\
\vec{n}_2 ~=~ \vec{n}_x ~-~ a \cdot D_\nu \vec{n}_x \,,\\
\vec{n}_3 ~=~ \vec{n}_x\,, \\
\vec{n}_4 ~=~ \vec{n}_x ~-~ a \cdot D_\mu \vec{n}_x\,,
\eea
where $\vec{n}_x$ is again given by (\ref{eigen-infinitesimal}) and $D$ is the covariant
derivative. It is straightforward then to evaluate $\Omega_x$
\beq
\label{omega}
\Omega_x ~=~ \frac{1}{2}\,a^2\, \left(\,
\vec{n} \cdot \left[ D_\mu \vec{n} \times D_\nu\vec{n}\right]
\right) ~+~ o( a^2 )\,.
\eeq
Therefore
\beq
\label{deficit}
\mathrm{'solid~~angle'~} ~=~  \frac{1}{4} \int_{S_C}\,
\vec{n} \cdot [ D_\mu \vec{n} \times D_\nu \vec{n}]\,\,d^2\sigma^{\mu\nu}\,.
\eeq

Unfortunately, there exists no simple expression for the third term, Eq.~(\ref{NAST-3}),
in the continuum limit. However, this is to be expected. Indeed, one
can readily convince oneself that the meaning of the 'perimeter contribution'
is to provide correct boundary conditions in Eq.~(\ref{NAST-3}).
In other words, the third term, Eq.~(\ref{NAST-3}), guarantees that the vector
field $\vec{n}(\sigma)\in S_\cC$ agrees with $\vec{n}(t)\in \cC$ on the boundary
$\delta S_\cC = \cC$.

Combining Eqs.~(\ref{dyn},\ref{deficit}) one formally reproduces Eq.~(\ref{Phase2})
confirming that (\ref{NAST-3}) is indeed the lattice formulation
of the non-Abelian Stokes theorem (\ref{Phase2}). However, this conclusion
relies heavily on Eq.~(\ref{omega}) which is only valid if $\vec{n}_x$, $x\in S_\cC$
is continuous across the plaquette boundaries. This suggests a natural way
to fix the relative sign of eigenstates on neighboring plaquettes
analogously to the continuum considerations above. Namely, we propose to fix
the particular distribution of eigenstates by the requirement that
\beq
\label{min}
R ~=~ \sum\limits_{x\in S_\cC} |\Omega_x| ~+~
\sum\limits_{x\in\cC} |\alpha_x|
\eeq
takes the minimal possible value (it is assumed, of course, that eigenvectors
at the boundary $\vec{n}(t) \in \cC$ are held fixed from the very beginning).
This prescription fixes completely and unambiguously all the states
$\vec{n}(\sigma)\in S_\cC$  provided that the functional $R$ has a unique minimum.
The uniqueness of the minimum of $R$ is a separate issue and we have
no analytical methods to investigate it. However, at least
numerically the minimum of (\ref{min}) might be approximated with high accuracy.

%====================================================================
\subsection*{Numerical simulations}
\noindent
In this Section we describe simple lattice experiments with Eq.~(\ref{NAST-3})
which we performed in pure SU(2) lattice gauge theory considered 
on $12^4$ lattice at $\beta=2.4$ using the standard Wilson action.

Since the decomposition (\ref{NAST-3}) is gauge invariant (see discussion in previous
Section) it is legitimate to ask what is the contribution of each term
into the Wilson loop expectation value
\beq
\label{full}
\langle\exp\{\,\, i \,\varphi_w\,\,\}\rangle ~\sim~ e^{-T V(R)}\,,
\eeq
where we have restricted ourselves to the consideration of rectangular $T\times R$,
$T \gg R$ loops only. Therefore the problem is to calculate
\beq
\label{dynamical}
\langle\exp\{\,\, i \sum\limits_{x\in S_\cC} \varphi_x\,\,\}\rangle ~\sim~ e^{-T V_{dyn}(R)}\,,
\eeq
\beq
\label{solid}
\langle\exp\{\,\, i \sum\limits_{x\in S_\cC} \Omega_x\,\,\}\rangle ~\sim~ e^{-T V_{solid}(R)}\,,
\eeq
\beq
\label{perim}
\langle\exp\{\,\, i \sum\limits_{x\in\cC} \alpha_x\,\,\}\rangle ~\sim~ e^{-T V_{perim}(R)}\,.
\eeq
Notice that $T, R$ dependence of the expectation values (\ref{dynamical}-\ref{perim})
is an ad hoc assumption which has to be checked separately. However, we have found
that (\ref{dynamical}-\ref{perim}) indeed accurately describe numerical data.

We calculated the expectation values (\ref{full}-\ref{perim}) on
50 statistically independent configurations using the spatial smearing
algorithm (see, e.g., Ref.~\cite{Bali:1994de} for details).
For each rectangular loop $\cC = \{T\times R\}$
Eqs.~(\ref{eigen0},\ref{chain}) were applied to construct the eigenstates
on $\cC$. The same procedure was used to build the eigenvectors 
$\{\vec{n}^{(i)}_p\}$, $i=0,...,3$ on each 'internal' plaquette $p\in S_\cC$
(only surfaces with minimal area were considered).
Finally, the functional (\ref{min}) was minimized with respect to the inversions
$\{\vec{n}^{(i)}_p\} \to \{-\vec{n}^{(i)}_p\}$, $p\in S_\cC$ using a variant of the
simulated annealing algorithm~\cite{Bali:1996dm} and keeping the boundary conditions
$\vec{n}\in \cC$ fixed.

\begin{figure}[ht]
\centerline{\psfig{file=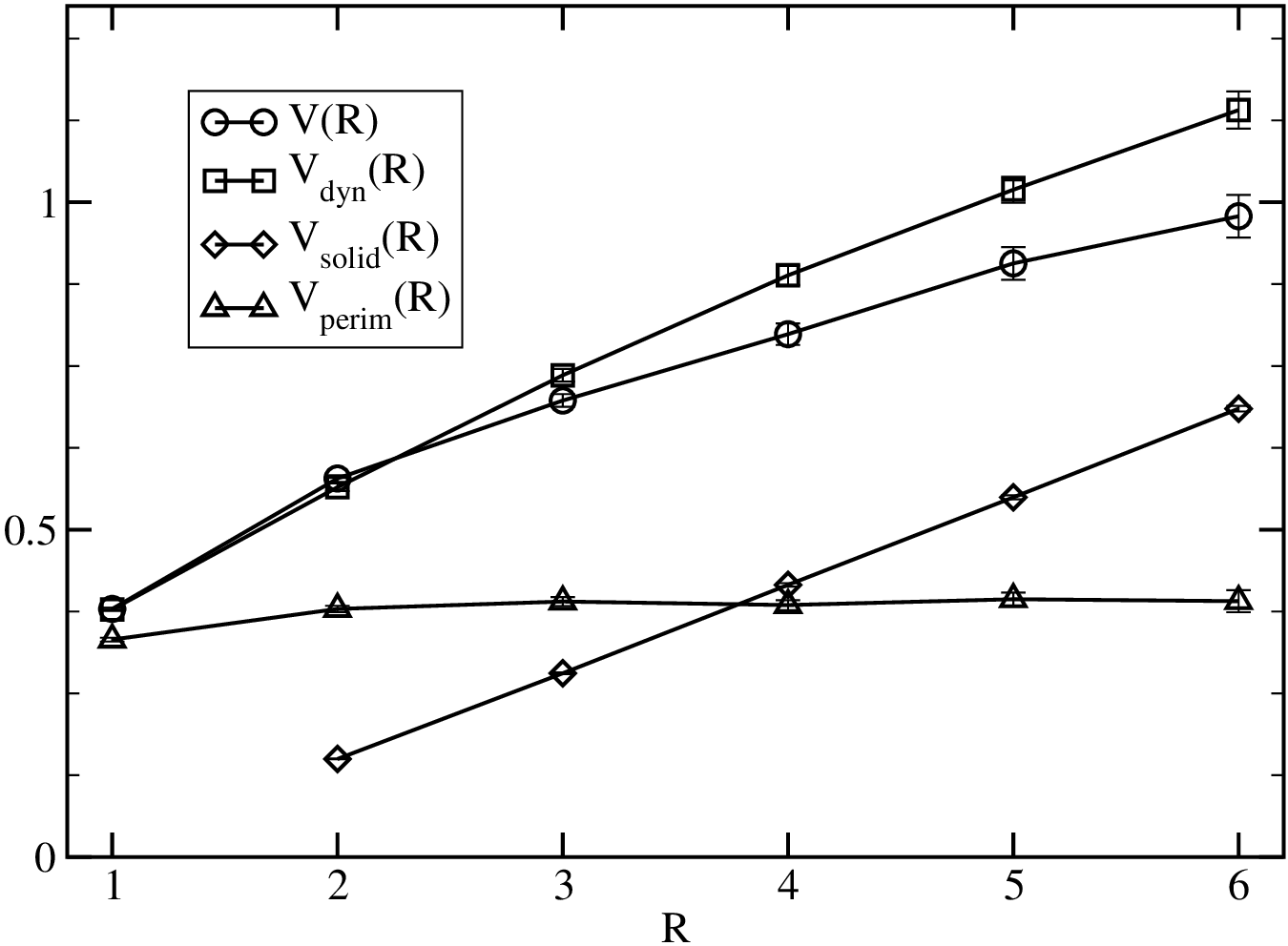,width=0.7\textwidth,silent=}}
\centerline{Fig.~3}
\end{figure}

The results of our simulations are presented on Fig.~3, where circles represent
the full heavy quark potential (\ref{full}), squares correspond the 'dynamical' part
(\ref{dynamical}) and finally diamonds and triangles stand for 'solid' (\ref{solid})
and 'perimeter' (\ref{perim}) contributions respectively. Note that the solid
curves on Fig.~3 are drawn to guide the eye.

There are few striking features of the expectation values (\ref{dynamical}-\ref{perim})
to be mentioned here. First of all, the 'perimeter' potential, Eq.~(\ref{perim}), 
turns out to be practically $R$-independent:
\beq
V_{perim}(R) ~\approx~  const\,,
\eeq
which might be an indication that the perimeter contribution drops out in the
expectation value of the full Wilson loop (\ref{full}). Secondly, both $V_{dyn}(R)$
and $V_{solid}(R)$ appear to be linear at large distances, $R \gtrsim 3$,
albeit with somewhat larger slope than the full potential
$V(R)\approx \sigma_{SU(2)} R$
\beq
\label{linear}
V_{dyn}(R) ~\approx~ \sigma_{dyn}\,R\,,
\qquad
V_{solid}(R) ~=~ \sigma_{solid}\,R\,,
\qquad
\frac{\sigma_{dyn}}{\sigma_{SU(2)}} \approx  \frac{\sigma_{solid}}{\sigma_{SU(2)}}
\approx 1.6\,.
\eeq
Although $V_{dyn}(R)$ deviates from the linear behavior at distances $R \lesssim 3$,
$V_{solid}(R)$ is rising strictly linear starting from the smallest possible
distance $R=2$. The existence of a linearly rising term in the heavy quark potential at short
distances has been widely discussed in the literature, see, e.g.,
Refs.~\cite{Vainshtein:1994ff}---\cite{Chetyrkin:1998yr}
and references therein.

Finally, we emphasize that the expectation value of the full Wilson loop
(\ref{full}) is not factorizable into the terms (\ref{dynamical}-\ref{perim}).
It is clearly seen from Fig.~3 that
\beq
V(R) ~\ne ~  V_{dyn}(R) ~+~ V_{solid}(R) ~+~ V_{perim}(R) ~+~ const
\eeq
and therefore there are various interference terms contributing to $V(R)$.
The point however is that the linear piece coming from $V_{solid}$
might survive at small distances since
the 'solid angle' contribution (\ref{deficit},\ref{solid},\ref{linear})
is formally not suppressed by the action even at very large $\beta$.

%====================================================================
\subsection*{Conclusions}
\noindent
We have derived a new version of the non-Abelian Stokes theorem
for the Wilson loop in the fundamental representation of $SU(2)$
gauge group. By considering the instantaneous color direction of the flux
piercing the loop we were able to avoid the path ordering in the conventional
definition of the Wilson loop operator. Moreover, this approach allows to represent
the phase angle of the Wilson loop as an ordinary one dimensional integral to which
the usual Stokes theorem applies. Furthermore, we were able to relate
the resulting surface integral with properties of non-Abelian gauge fields
on that surface.

Unfortunately, our formulation can hardly be called ``theorem'' because
it does not help to calculate the Wilson loop itself. However, this drawback
is not specific to this paper since other known variants of non-Abelian Stokes
theorem are also not much useful for Wilson loop calculation. At the same time
our construction is well suited for the numerical investigations. To achieve
this goal we have also derived the non-Abelian Stokes theorem on the lattice
and illustrated the origin and physical meaning of various terms contributing
to the trace of the Wilson loop.

%====================================================================
\subsection*{Acknowledgments.}
\noindent
We acknowledge thankfully fruitful discussions with T.~Suzuki and V.I.~Zakharov.
The work was supported by JSPS Fellowship No. P03024.

%==========================================================================================

\end{document}